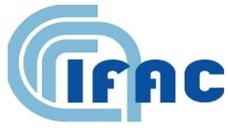

# The generation of concentrated energy of Bessel beams in an array of concentric rings

D. Mugnai[1]

[1]  *IFAC-CNR, Via Madonna del Piano 10, 50019 Sesto Fiorentino (FI), Italy*

The aim of this short note is to give information about the energy propagation of Bessel beams. Specifically, we refer to a Bessel beam which is generated by a source like that depicted in Fig. 1. In order to do this, we have to analyze that particular vectorial field that has the Bessel beam equation

$$u(\rho,\psi,z) = 2\pi A J_0(k_0\rho \sin \vartheta_0)\exp(ik_0 z \cos \vartheta_0)\exp(i\omega t) \qquad (1)$$

as its scalar approximation.[1]

For the source as depicted in Fig. 1 a detailed study of the propagation that uses a vectorial treatment has been performed [1]: the electric magnetic fields ($E$ and $H$, respectively) associated to a single ring are

$$\begin{aligned} E_x &= 2\pi\, e_0 e^{i\xi z}\left\{J_0(\eta\rho) + \tan^2\vartheta_0\left[\left(J_0(\eta\rho) - \frac{J_1(\eta\rho)}{\eta\rho}\right)\right.\right. \\ &\quad \left.\left. - \cos^2\psi\left(J_0(\eta\rho) - \frac{2J_1(\eta\rho)}{\eta\rho}\right)\right]\right\} \end{aligned} \qquad (2)$$

$$E_y = -2\pi\, e_0 e^{i\xi z}\left[\frac{\sin 2\psi}{2}\tan^2\vartheta_0\left(J_0(\eta\rho) - \frac{2J_1(\eta\rho)}{\eta\rho}\right)\right] \qquad (3)$$

$$E_z = -2\pi i\, e_0 e^{i\xi z}\left[\tan\vartheta_0 \cos\psi J_1(\eta\rho)\right], \qquad (4)$$

and

$$H_x = 0 \qquad (5)$$

$$H_y = \frac{2\pi}{Z} e_0 e^{i\xi z} \frac{1}{\cos\vartheta_0} J_0(\eta\rho) \qquad (6)$$

$$H_z = -i\frac{2\pi}{Z} e_0 e^{i\xi z} \sin\psi \frac{\sin\vartheta_0}{\cos^2\vartheta_0} J_1(\eta\rho), \qquad (7)$$

where $\xi = k\cos\vartheta_0$, $\eta = k\sin\vartheta_0$, and $J_1$ denotes the first-order Bessel function of first kind.

We note that the $E_x$ component, which represents the main contribution to the electric field, is practically coincident with Eq. (1) for for $\vartheta_0 \ll \pi/2$ ($r \ll f$), as in the present case.

From a knowledge of the electric and magnetic fields, we can now evaluate the mean density of the energy flux which is defined as being one half of the real part of the complex Poynting vector [2, 3], namely $S = \mathrm{Re}(E \times H^*)/2$. For the fields (2)-(4) and (5)-(7), we have

$$S_z = \frac{1}{2}\left(E_x H_y^*\right). \qquad (8)$$

The energy flux occurs only in the $z$-direction, and the energy propagates without any deformation.

When more than one beam impinges on the converging system ($C$ in Fig. 1), in order to have an increase in the energy (along the $z$-axis, see Fig. 1) each beam must be in phase with respect to the others. This condition is satisfied, for instance, if the optical path among the beams differs by one wavelength. It is then easy to find the geometric characteristics that the rings must have in order to supply localized energy.

Starting from the external ring-shaped aperture, let us denote with $R_1, R_2....R_n$ the optical paths, with $r_1, r_2......r_n$ the radii, and with $\vartheta_1, \vartheta_2.....\vartheta_n$ the axicon angles related to the n rings, respectively. We thus have

$$R_1 = \frac{f}{\cos \vartheta_1}$$

$$R_2 = R_1 - \lambda = \frac{f}{\cos \vartheta_2} \quad (9)$$

$$R_n = R_1 - (n-1)\lambda = \frac{f}{\cos \vartheta_n}$$

where $f$ is the focal length. Since the radius of the $n^{th}$ ring is $r_n = f \tan \vartheta_n$, it is immediately to find the radius of the rings such as the Bessel beams have the same phase. In Fig 2 we show the energy flux for a system consisting of 3 rings, suitably positioned following the relations (9).

---

[1] Equation (1) refers to a cylindrical coordinate system: $A$ is an amplitude factor, $\vartheta_0$ is the parameter which characterizes the aperture of the beam (Axicon angle), $J_0$ is the zero-order Bessel function of first kind [15], $k_0$ is the wavenumber in the vacuum. The field (1) is rotationally symmetric and thus independent of the angular coordinate $\psi$.

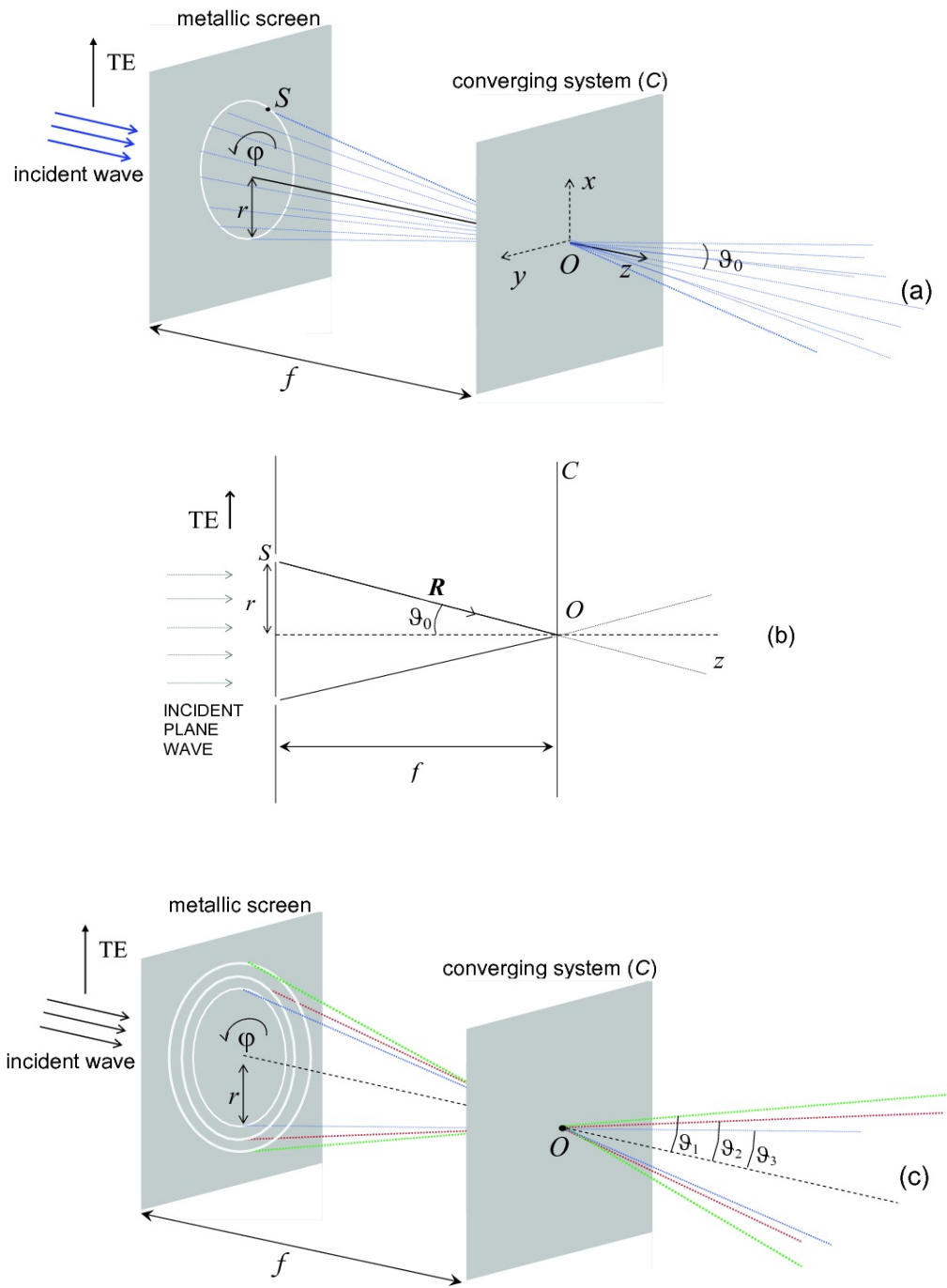

**Fig. 1** - (a) Scheme of the system considered for the vectorial treatment. From a theoretical point of view, the metallic screen and the converging system must be considered as being of infinite dimension. In practical applications this condition is satisfied if the dimensions of the metallic screen and of the converging system are much greater with respect to the wavelength. Moreover, also the condition $f \gg \lambda$ have to be satisfied; (b) details of the reference system; (c) scheme of the system in the presence of three rings.

As can be seen, the quantity of energy supplied by 3 rings is about nine times greater than the quantity due to a single ring (dotted line), as expected, due to the quadratic dependence of the Poynting vector with respect to the field.

We can also note that the energy maintains its localization, since the position of the first zero is unchanged. This effect is due to the small value of the ratio $\lambda/f$, which make the optical paths very close to each others. For higher values of the ratio $\lambda/f$, the axicon angles differ substantially one from the other and, consequently, the first zero of the Bessel function suffers a shift toward a higher value for $\rho$. In this situation, the amount of the energy flux is about the same, but the flux tends to lose its narrow localization.

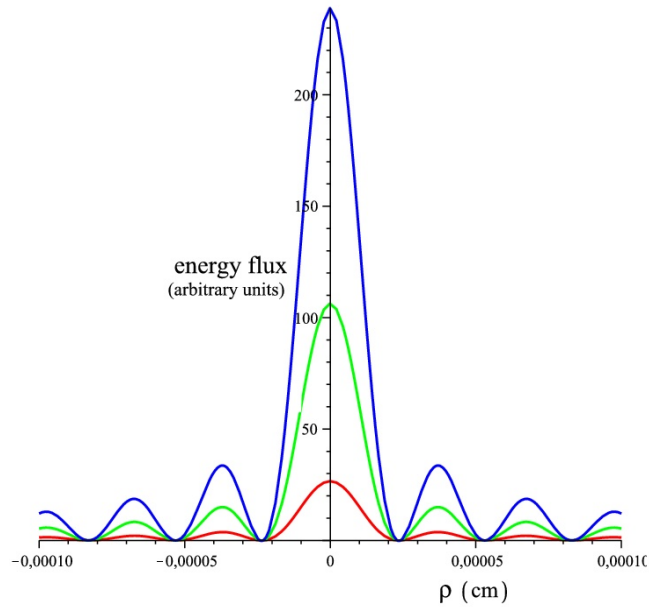

**Fig. 2** - Energy flux in the presence of one (red line), two (green line), and three ring-shaped apertures (blue line), as a function of $\rho$. Both the electric and magnetic fields are normalized to $e_0 \exp(iz\xi)$. Parameter values are: $f = 3 \times 10^{-2}$ cm, $\lambda = 3 \times 10^{-5}$ cm, $\vartheta_1 = 30°$.

**References**

[1] D. Mugnai and I. Mochi, Phys. Rev. E 73, 016606 (2006).
[2] J. D. Jackson, *Classical Electrodynamics*, Wiley, New York, 1999, Sec. 7.1.
[3] J.A. Stratton, *Electromagnetic Theory*, McGraw-Hill, New York, 1941, p. 342.